\documentstyle[twocolumn,aps]{revtex}

\input epsf

\begin{document} \title{Tuning p-wave interactions in an ultracold
Fermi gas of atoms} \author{C. A. Regal, C. Ticknor, J. L. Bohn,
and D. S. Jin \cite{adr1}} \address{JILA, National Institute of
Standards and Technology and Department of Physics, University of
Colorado, Boulder, CO 80309-0440} \date{\today} \maketitle

\begin{abstract} We have measured a p-wave Feshbach resonance
in a single-component, ultracold Fermi gas of $^{40}$K atoms. We
have used this resonance to enhance the normally suppressed p-wave
collision cross-section to values larger than the background
s-wave cross-section between $^{40}$K atoms in different
spin-states. In addition to the modification of two-body elastic
processes, the resonance dramatically enhances three-body
inelastic collisional loss.
\\~\\PACS number(s): 34.50.-s, 32.80.Pj, 05.30.Fk\\~\\
\end{abstract}

\narrowtext

The ultralow temperature regime accessible in atomic physics is
characterized by collision energies so low that centrifugal forces
can prevent scattering of atoms with nonzero relative angular
momentum. In this case the atoms collide predominantly via
s-partial waves ($l=0$).  Even this restricted regime has
demonstrated remarkably rich physics in atomic Bose-Einstein
condensates \cite{Stringari}.  However, the situation is
dramatically different for ultracold fermionic atoms. For
identical fermions the Pauli exclusion principle forbids s-wave
collisions.  This means the dominant interaction is via p-wave
collisions ($l=1$). These collisions are suppressed by centrifugal
effects as described by the Wigner threshold law
\cite{DeMarco3,Sadeghpour}, which demands that the elastic
scattering cross section diminishes with temperature as $\sigma
\propto T^2$. For this reason evaporative cooling of fermions to
ultralow temperatures can be achieved only in a mixture of
different spin states of the same atom \cite{DeMarco1,Granade} or
by sympathetic cooling with a different atom
\cite{Truscott,Schreck,Ketterle,Roati}.

However, p-wave collisions can become prominent in the presence of
a scattering resonance, such as a Feshbach resonance, which occurs
when the relative energy of an incident atom pair is nearly
degenerate with a quasi-bound molecular state. In ultracold
collisions Feshbach resonances are useful because the energy at
which they occur can be tuned by applying an external magnetic
field.  For this reason they have been exploited to measure atomic
interactions with great precision \cite{Roberts1,Chin} and to
influence the effective interaction between atoms \cite{Roberts2}.
In all previous cases Feshbach resonances have only been observed
in bosonic species \cite{Inouye,Courteille,Roberts3,Vuletic}, or
else between two distinct spin states of fermions
\cite{Loftus,Jochim,O'Hara,Dieckmann}.

In this Letter we report the first measurement of a magnetic-field
dependent Feshbach resonance in p-wave collisions between
ultracold fermionic atoms.  This resonance occurs between $^{40}$K
atoms in the $|F, m_F \rangle$ $=|9/2, -7/2 \rangle$ state as
predicted in Ref. \cite{Bohn} and results in p-wave interactions
that are nearly as strong as those from a nearby s-wave resonance
between the two distinct spin states $|9/2, -7/2\rangle$ and
$|9/2, -9/2\rangle$ \cite{Loftus}. We have measured and
characterized inelastic collisional losses near this p-wave
resonance, including loss at the nearby s-wave resonance.

\begin{figure} \begin{center} \epsfxsize=3.25
truein \epsfbox{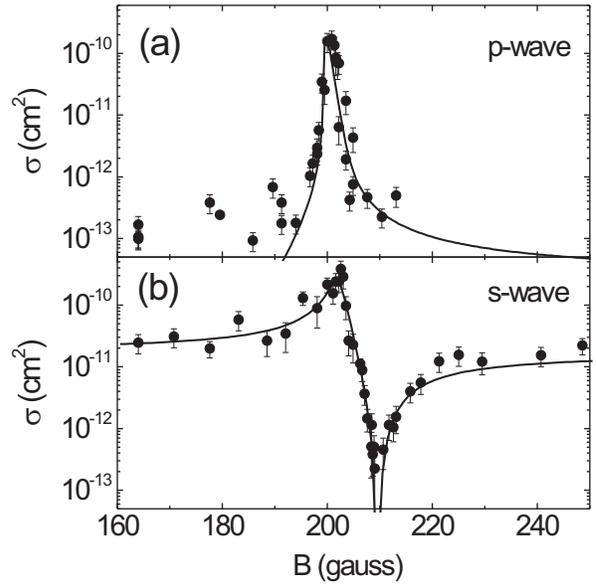} \end{center} \caption{Magnetic-field
Feshbach resonances between $^{40}$K atoms in the $|9/2,-7/2
\rangle$ state (a) and between atoms in the $|9/2,-9/2 \rangle$
and $|9/2,-7/2 \rangle$ states (b). The p-wave and s-wave data
were taken at temperatures of $3.2 \pm 0.7$ $\mu K$ and $4.4 \pm
0.9$ $\mu K$ respectively (both ${\rm T} \sim 2T_F$). The lines
are based upon our best-fit potassium potentials.}
\label{sandpwave}
\end{figure}

The experiments reported here employ previously developed
techniques for cooling and spin state manipulation of $^{40}$K.
Atoms in the $|9/2, 9/2\rangle$ and $|9/2, 7/2\rangle$ states are
first held in a magnetic trap and cooled by forced evaporation
\cite{DeMarco1}. The gas is then polarized in the $|9/2,
9/2\rangle$ state and loaded into a far-off resonance optical
dipole trap (FORT).  Adiabatic rapid passage is then used to
obtain the desired spin composition \cite{Loftus}. First, the
$|9/2, 9/2\rangle$ gas is completely transferred to the $|9/2,
-9/2\rangle$ state with a 10 ms rf frequency sweep across all ten
spin states at a field of $\sim 30$ G. To create a pure $|9/2,
-7/2 \rangle$ gas we then move to a higher field, $\sim 80$ G, and
sweep the magnetic field while applying rf at a fixed frequency to
drive the $|9/2, -9/2\rangle$ to $|9/2, -7/2\rangle$ transition.

A collision measurement is performed in the optical trap using
cross-dimensional rethermalization as described in Ref.
\cite{Loftus}. Our optical trap is characterized by transverse
frequencies of $\nu_y \approx 1$ kHz and $\nu_x$ = 1.7 $\nu_y$ and
an axial frequency of $\nu_z$ = $\nu_y$/80. For high collision
rates we use rethermalization between the two transverse
directions to ensure that the collision time, not the trap period,
defines the rethermalization time $\tau$. We extract the elastic
scattering cross section $\sigma$ via the relation ${1 \over \tau}
= {2 \over \alpha} \langle n \rangle \sigma v$, where the mean
relative speed is given by $v=4\sqrt{k_BT / \pi m}$ and the number
density is given by $\langle n \rangle =\frac{1}{N}
\int\,n(\vec{r})^2 \,d^3\vec{r}$, where $N$ is the total number of
atoms. The constant $\alpha = 4.1$ is the calculated average
number of binary p-wave collisions per atom required for
thermalization \cite{DeMarco3}.

Cross sections for p-wave collisions measured in this way are
shown in Fig.~\ref{sandpwave}(a) as a function of the magnetic
field.  The dominant feature of this resonance is a peak in
$\sigma$ that rises over 3 orders of magnitude above the small
background cross section.  To the low field side of this peak the
effective p-wave interactions are expected to be repulsive, while
to the high field side they should be attractive. The zero
crossing of the resonance, which is predicted to be $\sim 20$ G
below the peak, is not observed experimentally. Instead, away from
the resonance peak we measure a background value of $\sim
10^{-13}$ cm$^2$, consistent with a spin state impurity of $<
0.2\%$.

For comparison we also show in Fig.~\ref{sandpwave}(b) the s-wave
resonance observed between $|9/2, -7/2\rangle$ and $|9/2,
-9/2\rangle$ atoms. This plot includes data from Ref.
\cite{Loftus} as well as additional measurements near the peak of
the resonance. The presence of non-negligible off-resonant
scattering in the s-wave case makes the characteristic asymmetry
of the Fano lineshape more apparent for this resonance than for
the p-wave resonance \cite{Fano}. Note that, despite the
differences in the background p-wave and s-wave cross sections,
the strengths of these partial waves become comparable on
resonance.

For these data, the magnetic field values B were calibrated using
rf-driven Zeeman transitions and have a systematic uncertainty of
$\pm 0.1 \%$. The error bars in $\sigma$ are dominated by
measurement uncertainties in $\tau$.  In addition, $\sigma$ has an
overall systematic uncertainty of $\pm 50 \%$ that comes from
measurements of $N$.  The curves in Fig.~\ref{sandpwave} are
best-fit theory calculations using the standard close-coupling
theory of ultracold collisions and are appropriately thermally
averaged \cite{pwave}. The relative position of the s-wave and
p-wave resonances depends sensitively on the $C_6$ van der Waals
coefficient.  This arises from the role that the van der Waals
potential plays in setting the centrifugal barrier height for
p-wave collisions. Measuring the separation in magnetic field of
the two resonances therefore enables us to determine a value of
$C_6 = 3902 \pm 15$ atomic units, in good agreement with the
predicted value of $C_6 = 3897 \pm 15$ \cite{derevianko}. The
values for $a_t$ and $a_s$ obtained from the simultaneous fit are
consistent with the values quoted in Ref. \cite{Loftus}.

One striking feature of Fig.~\ref{sandpwave} is the close
proximity of the p-wave resonance to the previously reported
s-wave resonance in $^{40}$K.  At zero temperature, the two
resonances are separated in field by only 3.2 $\pm$ 0.8 G. This
result is apparently purely coincidental. By artificially varying
either $C_6$, $a_s$, or $a_t$ for potassium we have calculated
that the resonances both shift to different values of the field.
For example, if the singlet scattering length had been 165 $a_0$
instead of 104 $a_0$, the resonances would have been separated by
$\sim 70$ G.

\begin{figure} \begin{center} \epsfxsize=3.25 truein
\epsfbox{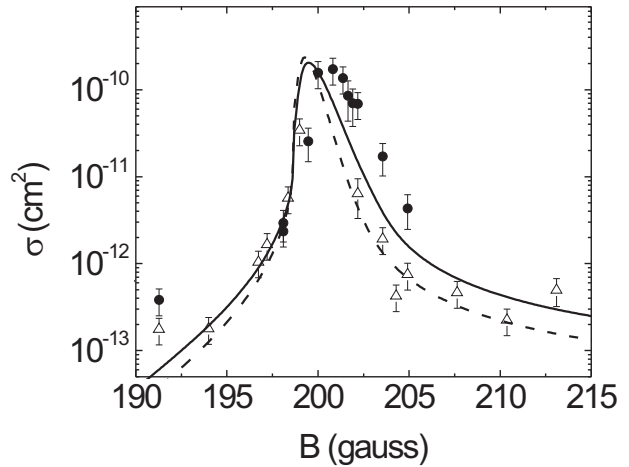} \end{center} \caption{Temperature dependence
of the p-wave elastic collision cross section.  The two
temperatures sets are $3.7 \pm 0.5$ $\mu K$ (${\Large \bullet}$)
and $2.7 \pm 0.3$ $\mu K$ ($\triangle$).} \label{pwave}
\end{figure}

To view the temperature dependence of the p-wave resonance, we
plot in Fig.~\ref{pwave} the elastic scattering cross section for
$|9/2,-7/2\rangle$ collisions at two temperatures. For this plot
the data from Fig.~\ref{sandpwave} are divided into two
temperature sets and compared to thermally averaged theoretical
results based on our best-fit potassium potentials. We observe
that the resonance lineshape in Fig.~\ref{pwave} displays an
asymmetry characteristic of near-threshold resonances, as is seen
in photoassociation spectroscopy \cite{Napolitano}. The onset of
the resonance at 198.4 $\pm$ 0.5 G is sharp and relatively
insensitive to temperature; this is the field value at which a
high-lying molecular bound state appears as a low-energy
resonance, and can be accessed at any temperature.  In contrast,
the high field tail of the asymmetric peak is a strong function of
temperature.

In addition to elastic scattering, inelastic collision rates can
be altered at Feshbach resonances.  For the combinations of spin
states we are studying, spin-exchange collisions are energetically
forbidden at these temperatures and magnetic field strengths.
Thus, for the p-wave resonance the dominant two-body inelastic
collision process is the reaction $|9/2,-7/2 \rangle + |9/2,-7/2
\rangle$ $\rightarrow |9/2,-7/2 \rangle + |9/2,-9/2 \rangle$,
driven by a comparatively weak magnetic dipole interaction. On the
other hand, for the s-wave resonance, the $|9/2,-7/2\rangle$ and
$|9/2, -9/2 \rangle$ states are distinguished by being the two
lowest-energy states of the $^{40}$K atom at the magnetic fields
in this experiment; they are therefore immune to any two-body
losses associated with the s-wave resonance. Further, by symmetry
there can be no p-wave Feshbach resonance between atoms in these
two spin states of $^{40}$K.

Significant trap loss can also arise from three-body collisions,
in which two atoms recombine into a bound molecular state, while
the third atom carries away the binding energy. For bosons these
collisions can cause a significant limitation to trap lifetimes
near a Feshbach resonance \cite{Inouye,Roberts1}. For ultracold
fermions three-body interactions can be suppressed by Fermi
statistics \cite{Esry}.  However, this statistical suppression
does not exist at a scattering resonance, and thus three-body loss
is expected near the peaks of our Feshbach resonances.

\begin{figure} \begin{center} \epsfxsize=3.25  truein
\epsfbox{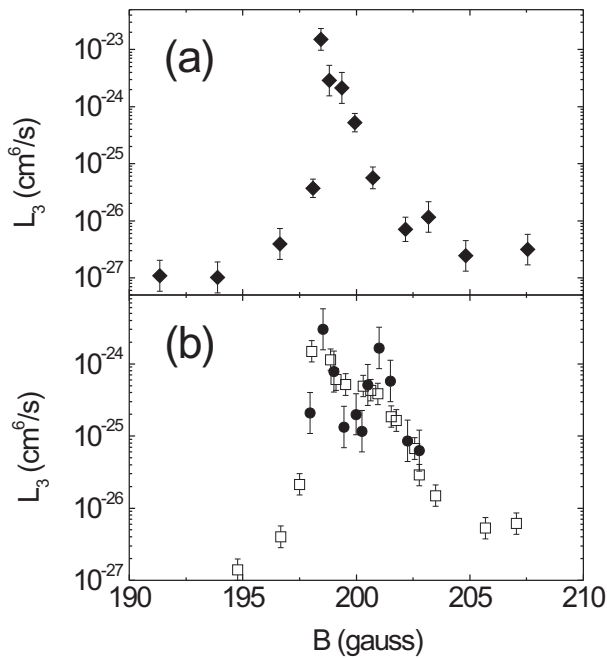} \end{center} \caption{Resonant three-body
loss for a gas of atoms in the (a) $|9/2,-7/2 \rangle$ state and
(b) $|9/2,-7/2 \rangle$ and $|9/2,-9/2 \rangle$ states.  The data
in (a) were taken at a temperature of $2.4 \pm 0.8$ $\mu K$ while
the data in (b) were taken at $1.25 \pm 0.10$ $\mu K$ (${\Large
\bullet}$) and $2.8 \pm 0.4$ $\mu K$ ($\Box$). The error bars are
determined from scatter among repeated loss measurements.}
\label{loss}
\end{figure}

We have probed inelastic collisions near the p-wave resonance by
tracking depletion of trapped atoms. Figure~\ref{loss}(a) presents
the inelastic collisional loss near the p-wave resonance for a gas
of $|9/2,-7/2 \rangle$ atoms. Here we assume the loss is dominated
by three-body processes and characterize the loss in terms of a
three-body atom loss rate $L_3$. Due to our uncertainty in
measured N ($\pm 50 \%$), all quoted values for $L_3$ have a
systematic uncertainty of a factor of three. Near the resonance
the loss rate varies by orders of magnitude, just as the elastic
rate does, with rate constants comparable to those seen for
bosons\cite{Roberts4,Stenger}.  The loss rate also shows the
characteristic asymmetry arising from threshold effects, as seen
in the elastic two-body cross section.

Because this p-wave resonance is likely to affect inelastic
collision processes at the nearby s-wave resonance, we have also
measured atom loss rates for a gas containing an equal mixture of
$|9/2,-9/2\rangle$ and $|9/2,-7/2\rangle$ atoms. Here we examined
the atom loss across the peaks of both the p-wave and s-wave
resonances, whose zero temperature resonance positions are located
at 198.4 $\pm$ 0.5 G and 201.6 $\pm$ 0.6 G respectively. As shown
in Fig.~\ref{loss}(b), the loss at a temperature of $2.8 \pm 0.4$
$\mu K$ is a broad peak that encompasses both resonances. At a
lower temperature of $1.25 \pm 0.10$ $\mu K$ the loss appears as
two distinct peaks, one at the same magnetic field as the p-wave
loss peak and the other near the peak of the s-wave resonance.

In general, three-body loss processes in a Fermi gas are
influenced by Fermi exchange symmetry. With two distinct spin
states present, a three-body collision may involve either a pair
of atoms in one spin state and the third atom in the other, or
else all three atoms in the same spin state. Away from resonance,
this difference combined with Fermi exchange symmetry alters the
threshold behavior of three-body collisions \cite{Esry}.  On
resonance, the influence of this dynamics remains to be explored.

\begin{figure}[top] \begin{center} \epsfxsize=3.25 truein
\epsfbox{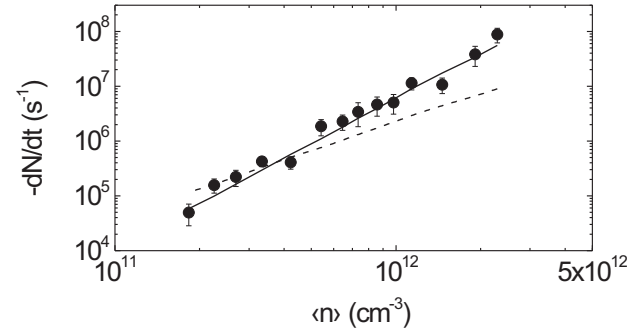} \end{center} \caption{Density dependence of
$|9/2,-7/2 \rangle$ atom loss at B = 198.8 G.  The lines are
results of best-fit curves that assume either exclusively two-body
loss (dashed line) or exclusively three-body loss (solid line).}
\label{lossproof}
\end{figure}

\begin{figure} \begin{center} \epsfxsize=3.25 truein
\epsfbox{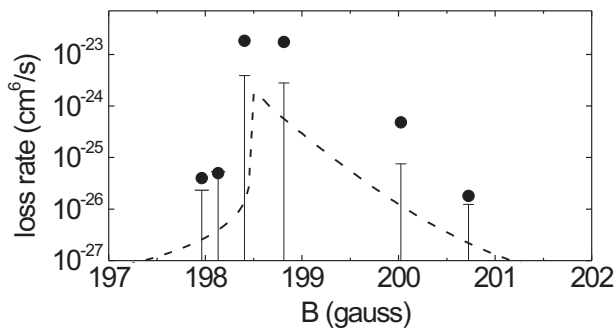} \end{center} \caption{Comparison of
three-body and two-body loss of $|9/2,-7/2 \rangle$ atoms across
the p-wave resonance at a temperature of $2.7 \pm 0.7$ $\mu K$.
Data shown are $L_3$ (${\Large \bullet}$) and an experimental
upper limit on the value of $L_2$ (solid lines). For comparison to
$L_3$, $L_2$ is divided by a typical density of $\langle n \rangle
= 3 \times 10^{12}$ cm$^{-3}$. The dashed line is the result of a
close-coupling calculation of the two-body inelastic atom loss
rate.} \label{K2K3}
\end{figure}

We have also performed a study of the density dependence of the
observed loss at the p-wave resonance through measurements of loss
of atoms in time as the density of the gas changed by an order of
magnitude. The loss trends were analyzed with the following rate
equation \cite{Roberts4}.
\begin{equation} \frac{dN}{dt} = -L_2 \langle n \rangle N - L_3
\langle n^2 \rangle N - \frac{N}{\tau_0}
\end{equation} Here, $\langle n^2 \rangle = \frac{1}{N} \int\,n(\vec{r})^3
\,d^3\vec{r}$, and $\tau_0 = 32$ seconds describes the exponential
lifetime of atoms in our optical trap. The time evolution of the
size of the gas is included in the density terms. For a field of
198.8 G, we plot in Fig.~\ref{lossproof} the number loss $dN/dt$
as a function of the atomic density $\langle n \rangle$. The solid
and dashed lines in Fig.~\ref{lossproof} are best fits assuming
exclusively three-body loss ($L_2 = 0$) or two-body loss ($L_3 =
0$) respectively. The data at this field are clearly consistent
with three-body, but not two-body, losses. Figure~\ref{K2K3}
presents the results of this procedure at various field values
across the p-wave resonance. Here we compare the two-body and
three-body loss at a density typical of the initial density of our
cloud, $\langle n \rangle = 3 \times 10^{12}$ cm$^{-3}$. At the
peak of the Feshbach resonance the three-body loss clearly
dominates over the limiting value of $L_2$. For comparison we also
plot the theoretically expected two-body loss rate, which lies
within the measured upper limit for this rate. A similar analysis
of the loss of $|9/2, -7/2 \rangle$ and $|9/2, -9/2 \rangle$ atoms
near the s-wave peak shows that this loss process is also
predominantly three-body at our typical densities.

In conclusion, we have observed a p-wave Feshbach resonance for
ultracold fermions and characterized both the elastic and
inelastic processes near this resonance in $^{40}$K.  The observed
elastic cross sections for the combination of the p-wave and
nearby s-wave resonances constrain $C_6$ for potassium. Further,
the modification of the elastic cross sections provided by these
resonances yields ultracold fermions with tunable interactions. In
particular, the availability of both p-wave and s-wave resonance
offers the opportunity to study strong interactions between both
identical and non-identical fermions. Among the most interesting
prospects for such strong interactions is driving Cooper pairing
of fermionic atoms \cite{Stoof,Holland,Timmermans,nature}.
However, three-body interactions result in dramatic losses at
these $^{40}$K resonances. Thus, understanding on-resonance
threshold behavior of the three-body collisions involved will be
critical to utilizing these resonances to achieve a strongly
interacting Fermi gas of atoms.

This work is supported by NSF, ONR, and NIST.  C. A. Regal
acknowledges support from the Hertz Foundation.  We also
acknowledge useful conversations with C. H. Greene, B. D. Esry,
and H. Suno, and we thank M. L. Olsen for contributions.


\begin{thebibliography}{10}
\bibitem{pwave}
For the p-wave we fit only points with $\sigma > 10^{-12}$ cm$^2$.
\end{thebibliography}

\end{document}